\documentclass{CUP-JNL-DCE}

\usepackage{latexsym}
\usepackage{graphicx}
\usepackage{multicol,multirow}
\usepackage{amsmath,amssymb,amsfonts}
\usepackage{mathrsfs}
\usepackage{amsthm}
\usepackage{apacite}
\usepackage{rotating}
\usepackage{appendix}
\usepackage[authoryear]{natbib}
\usepackage[capitalise]{cleveref}
\usepackage{breakcites}
\usepackage{lscape}

\articletype{RESEARCH ARTICLE}
\jname{Data-Centric Engineering}
\jyear{2020}

\graphicspath{ {./figures/} }

\begin{document}

\begin{Frontmatter}

\title[Context-specific volume-delay curves by combining crowd-sourced traffic data with Automated Traffic Counters (ATC): a case study for London]
{Context-specific volume-delay curves by combining crowd-sourced traffic data with Automated Traffic Counters (ATC): a case study for London}

\author*[1]{Gerard Casey}\email{gerard.casey@arup.com	}
\author[2]{Bingyu Zhao}\email{bz247@berkeley.edu}
\author[3]{Krishna Kumar}\email{krishnak@utexas.edu}
\author[2]{Kenichi Soga}

\authormark{Gerard Casey \textit{et al.}}

\address*[1]{\orgname{Arup}, \orgaddress{\state{London},  \country{UK}}}
\address[2]{\orgdiv{Department of Civil and Environmental Engineering}, \orgname{University of California, Berkeley}, \orgaddress{ \country{USA}}}
\address[3]{\orgdiv{Department of Civil, Architectural and Environmental Engineering}, \orgname{University of Texas at Austin}, \orgaddress{ \country{USA}}}


\abstract{Traffic congestion across the world has reached chronic levels. Despite many technological disruptions, one of the most fundamental and widely used functions within traffic modelling, the volume delay function, has seen little in the way of change since it was developed in the 1960's. Traditionally macroscopic methods have been employed to relate traffic volume to vehicular journey time. The general nature of these functions enables their ease of use and gives widespread applicability. However, they lack the ability to consider individual road characteristics (i.e. geometry, presence of traffic furniture, road quality and surrounding environment). This research investigates the feasibility to reconstruct the model using two different data sources, namely the traffic speed from Google Maps' Directions Application Programming Interface (API) and traffic volume data from automated traffic counters (ATC). Google's traffic speed data are crowd-sourced from the smartphone Global Positioning System (GPS) of road users, able to reflect real-time, context-specific traffic condition of a road. On the other hand, the ATCs enable the harvesting of the vehicle volume data over equally fine temporal resolutions (hourly or less). By combining them for different road types in London, new context-specific volume-delay functions can be generated. This method shows promise in selected locations with the generation of robust functions. In other locations it highlights the need to better understand other influencing factors, such as the presence of on road parking or weather events.}

\begin{policy}[Impact Statement]
Volume delay curves are widely used in traffic analysis. They form a critical part of the traffic assignment stage in the four-step modelling approach. We propose a novel data-driven approach to accurately represent the traffic behaviour under congestion using emerging data sources for a better understanding of real-world context-specific behaviour.
\end{policy}

\end{Frontmatter}
\section{Introduction}\label{introduction}
When the vehicular demand for a road exceeds its free-flow threshold, a journey time-delay is incurred, resulting in congestion. Historically in traffic studies and planning, the relationship between traffic volume and time-delay has been simplified to macroscopic principles ignoring the behaviour of individual vehicles~\citep{van2015genealogy}. The most widely adopted functional representation of such a macroscopic volume-delay relationship is the Bureau of Public Roads (BPR) relationship proposed in the 1960s~\citep{us1964traffic}. Since then, local traffic agencies worldwide have calibrated the BPR curve coefficients to suit the local needs, such as in~\citet{SUH1990177, kurth1996implementation,irawan2010implementation, mtoi2014calibration, kucharski2017estimating}. However, the calibration of BPR coefficients to local roads requires an extensive set of volume and delay data, and gathering such data (e.g., through surveys) is often a challenge. Hence, traffic engineers often end up using a generic set of coefficients to cover a wide range of roads (for example, one set of coefficients for all roads in the city, without considering the presence of road furniture or on-street parking that could affect the road capacity) and do not updated these coefficients to the changing demands of the city. Nevertheless, real-world conditions exhibit an enormous variability in the volume-delay characteristics, which is nearly impossible to account for entirely and certainly insufficient when representing this volume-delay relationship using a generic and static function for the entire city. Therefore, it is critical to understand the errors in volume-delay prediction based on real-world data.

This research aims to provide an efficient data-driven approach to calibrating the volume-delay curves by incorporating emerging data sources, mainly crowd-sourced travel time information from location and routing service providers, such as Google Maps~\citep{google_maps}. This type of calibration has never been done before to our knowledge. Specifically, this paper investigates the integration of two disparate data sources for this purpose: (1) the traffic speed data from Google Maps, and (2) the traffic count data from the Automated Traffic Counter (ATC) system in Greater London. The reason to combine the two data sources is two-fold: (1) first of all, many traffic detectors in cities are single loop detectors, which do not offer accurate speed/delay measurement~\citep{wang2003can}; (2) secondly, even if there are speed sensors (e.g., double loop detectors or radars), they often only measure point speeds, which do not reflect the delay experienced on the road network, especially in urban settings~\citep{wang2005empirical, zhang2015accuracy}. In this study, we assess the feasibility and performance of obtaining site-specific volume-delay information based on these new data sources. 

The paper is structured in the following way. Section 2 discusses relevant literature, including various functional forms of the volume-delay relationships, as well as some applications of such relationships in transport research and planning. Next, in Sections 3 and 4, we introduce the two data sources, namely the hourly traffic volume from the ATCs and the traffic speed from Google Maps. These data are used for estimating the road characteristics (free-flow travel time and capacity) and calibrating the volume-delay curve coefficients. Sections 5 and 6 explain the process of data cleaning and model building. In particular, three models are tested, including one base model with default volume-delay parameter values and two data-informed models with partial or full set of parameters calibrated from the real-world data. The three models are compared in terms of their performances in quantifying the level of delays at different traffic congestion levels. Section 7 offers an extensive discussion on the potential factors that could cause the variability observed in the volume-delay relationship, as well as the future prospects to adopt the proposed method at larger scales. Using fine-resolution data in the form of ATCs and aggregated device-based location-informed journey times on a range of roads, we demonstrate the capability of the new data-driven approach for efficiently capturing the volume-delay characteristics of roads in selected roads in Greater London.

\section{Literature review}
Volume-delay functions, as the name indicates, relate two fundamental traffic parameters using non-linear mathematical expressions. The independent parameter, volume, expresses the level of traffic demand and the dependent parameter, delay, indicates the deterioration in the traffic speed as the demand increases. The actual trend between volume and delay is a property of the road and is related to factors such as the speed limit, width, geometry, and the presence of road furniture. Even though the volume-delay relationships are highly context-specific, there exist some widely accepted functional forms to model them. The most widely used function is the BPR curve, which was developed in the 1950s for uncongested freeways in the US. Its simple mathematical form and minimal input requirements are attributed to its widespread adoption~\citep{skabardonis1997improved}. The travel time $t$ on a road link computed using the BPR function has the form:

\begin{equation}
	t = t_0 \times (1 +  \alpha(\frac{v}{v_c})^\beta )
	\label{eq:BPR}
\end{equation}

\noindent
where $t_0$ is the time required to traverse the road link at free-flow speed; $v_c$ is the road capacity (vehicles per unit time); $\alpha$ and $\beta$ are calibration coefficients; $v$ is the traffic volume to be modelled. The function is sometimes also expressed in terms of the vehicle speed $u$, which can be obtained from $t$ and the road length $l$.

Since the BPR function was created by fitting a polynomial equation to uncongested freeway data from the 1950s in the US, it does not reflect the current operating conditions of the road network~\citep{skabardonis1997improved}. As a result, many different organisations have adapted the BPR curve with various local empirical and simulated data to suit the local road conditions better~\citep{irawan2010implementation, mtoi2014calibration}. The work of~\citet{kurth1996implementation} focuses on obtaining refined, free-flow time and capacity ($t_0$ and $v_c$) for each road based on guidelines from the Highway Capacity Manual (HCM, 1994 version), rather than getting $t_0$ and $v_c$ from traditional lookup tables with only a few categories. Their approach is found to produce more accurate traffic speed estimations. However, it requires time-consuming identification of road characteristics (e.g., road grade, vehicle mix and land use) and is still limited by the available adjustment factors that the HCM procedure can take into account.

\citet{kucharski2017estimating} estimated both the calibration coefficients $\alpha$, $\beta$ and the road characteristics $t_0$, $v_c$ together based on loop detector data. They suggested transforming the volume-delay relationship to speed-density relationship for regression, as the latter remains monotonic in congested cases. However, the analysis lacks rigour due to employing linear R-squared metric to quantify errors in a non-linear model and inconsistency in calculating the density. In the case of London, Transport for London (TfL) has calibrated the BPR function using observed traffic counts and defined $\alpha=1.0$ and $\beta=2.0$ for the area~\citep{tfl_modelling_guidelines}. In general, these calibrated volume-delay functions fit the local observations better.

Apart from the BPR function, several volume-delay relationships have been proposed over the years, as summarised in~\citet{mtoi2014calibration}.~\citet{davidson1966flow} proposed a general-purpose travel-time formula in 1966 and this method has undergone numerous modifications since it was first proposed~\citep{mtoi2014calibration}. It has exhibited a closer match to actual volume counts and has a more robust theoretical base than the BPR~\citep{rose1989estimating}. Among the modifications of the Davidson function~\citep{tisato1991suggestions,akcelik1991travel}, the Ak\c{c}elik function is the most widely used. The Ak\c{c}elik method is a time-dependent modification of the Davidson model, which uses the coordinate transformation technique in an attempt to overcome the conceptual and calibration issues with the Davidson method~\citep{akcelik1991travel}. This function showed good results for certain road types, tolls roads and signalized arterials~\citep{mtoi2014calibration}.~\citet{spiess1990conical} proposed the conical method, which attempts to overcome some of the limitations of BPR at both the upper and lower bounds by employing hyperbolic conical sections, while maintaining a similar form to the BPR. The similarity to BPR enables a direct transfer of parameters. These alternative formulations have also been adopted in practice and research.

Volume-delay functions are typically employed in utility estimation (e.g., time cost) for static or semi-dynamic traffic assignments, and informing route choices for agent-based modeling~\citep{ccolak2016understanding, SUH1990177}. The differentiable form and convex nature of volume-delay functions make them an ideal candidate for optimisation-based traffic assignment, such as the assignment that satisfies the Wardrop's equilibrium~\citep{lien2016wardrop}. Despite their extensive use in research and practice, the volume-delay functions have several limitations. For example, it is possible to obtain traffic volume-to-capacity ratio much higher than one, which is unrealistic~\citep{nie2004models, chiu2011dynamic}. Closely associated is the problem that the volume-delay functions can only model the hypocritical section of the traffic fundamental diagram (monotonic increase in the flow on a link with travel time and density), but not the hypercritical section (when a road is congested to a certain level, the flow will decrease despite an increase in density and travel time). These limitations mean that volume-delay functions cannot model traffic phenomena such as spillbacks, wave propagation, and gridlocks, which would require the use of a dynamic model~\citep{lo2005road, chiu2011dynamic}. These concerns are also reflected in the UK Department for Transport (DfT) Transport Analysis Guidance (TAG), which recommended to model junction delays explicitly especially for congested urban roads~\citep{DfT_webtag}. Despite these shortcomings, volume-delay functions are a useful tool for regional-scale simulations and analysis requiring faster computations. Carefully calibrated volume-relationships frequently show good match with the real-world data~\citep{kurth1996implementation, irawan2010implementation}.

In order to effectively use volume-delay functions in regional-scale analyses, it is crucial to maintain an up-to-date coefficients specific to local regions, which has proven to be a nontrivial task. Past studies have identified the need to inform these functions with empirical and context-specific data~\citep{rose1989estimating,spiess1990conical}, but also recognise the difficulty and cost associated with collecting such empirical data as being prohibitive~\citep{rose1989estimating}. There have been studies that incorporate different forms of field sensor data for such calibration~\citep{mtoi2014calibration, neuhold2014volume, kucharski2017estimating}. Nevertheless, the data used in these studies have been generated specifically for that application and requires specific hardware and software for use.

Recent innovations in the information and communication technologies have led to an increase in the adoption of real-time crowd-sourced data feeds in transport modelling. This includes applications of location data for emissions estimations~\citep{hirschmann2010new}, building origin and destination matrices~\citep{toole2015path} and general urban traffic management applications~\citep{artikis2014heterogeneous}. This study investigates the use of a novel real-time crowd-sourced data feeds that have wider spatial coverage and are not generated specifically for estimating volume-delay functions. These sources could be used to consider some of these previously ignored characteristics and create temporally and spatially dynamic volume, speed and saturation relationships. Such data sources can harvest data at a finer resolution over a longer (even indefinite) period of time giving a far greater understanding of the temporal variations and trends exhibited on road infrastructure.

This work attempts to combine new data sources in order to create functions that do not require a large range of survey inputs. The general and transparent methodology of harvesting crowd-sourced data enables its easy deployment to multiple sites. As a result, highly localised relationships can be obtained for a diverse range of road links, reflecting individual characteristics of the road (i.e. geometry, the presence of traffic furniture, road quality and the surrounding land use). Such varying characteristics can result in very different vehicular behaviour on roads that may be considered similar by traditional approaches. 

\section{Traffic volume inputs: ATCs}

Two sources of data inputs are utilised to calibrate the volume-delay characteristics localised to the road link level: the link-level traffic counts (volume) from the ATCs and the link-level travel time (delay) from Google Maps' real-time information. Data from these two sources were harvested over the same time period (late February to mid March, 2016) and then paired according to the time of collection to create volume-delay observations. These observations will be used later in this paper to calibrate the context-specific volume-delay functions for road links where observations are available. In this and the next section, the two data sources will be introduce.

ATCs are magnetic induction loops embedded under the road surface. The passing of a vehicle results in an electromagnetic signal. The ATCs in Greater London count every vehicle which passes over the inductions loop. The data used here was harvested over a period of three weeks from the 27th February to the 21st March 2016.

\begin{figure}[htbp!] 
	\centering    
	\includegraphics[width=0.75\textwidth]{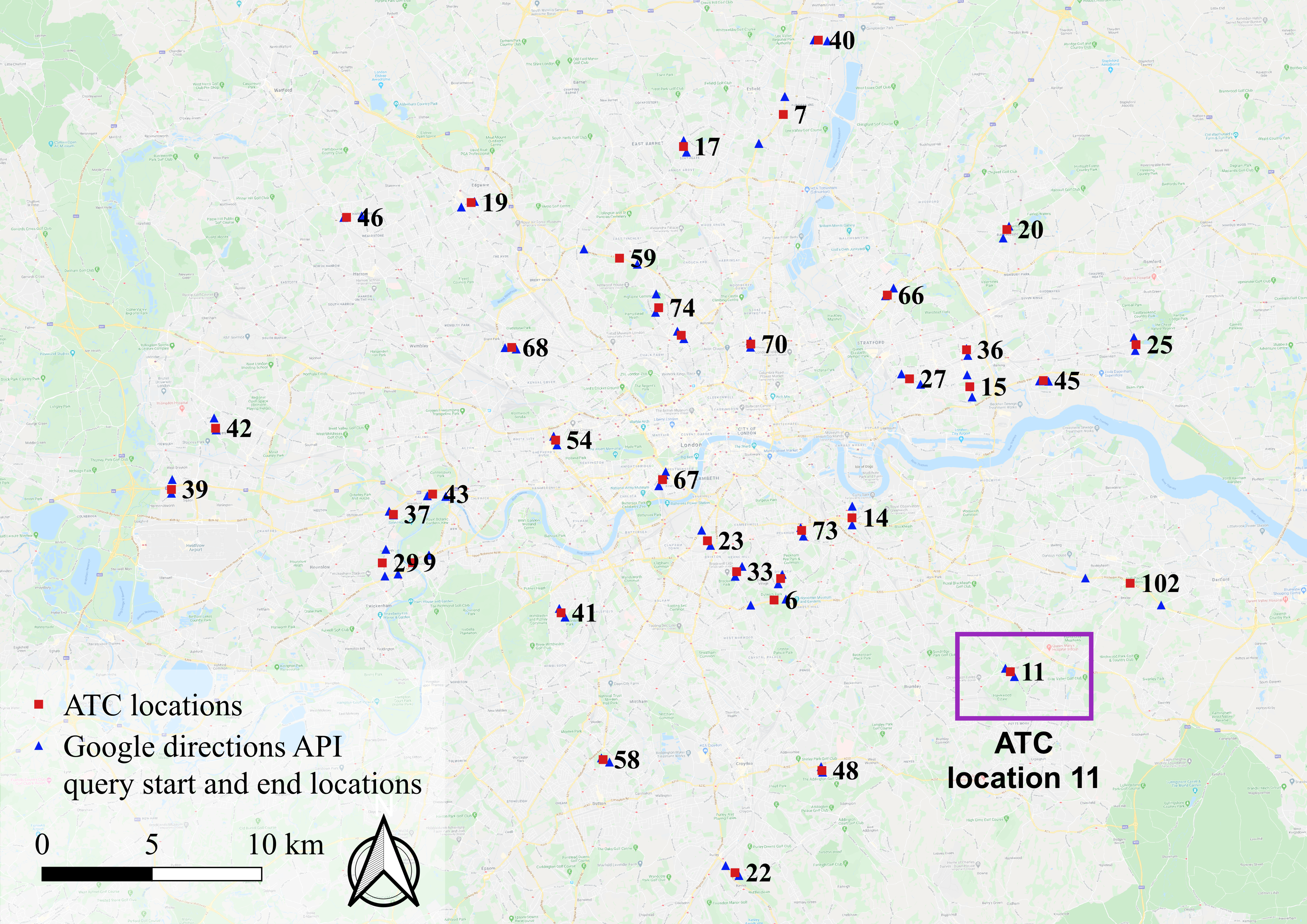}
	\caption[Map of ATC locations in Greater London]{ATC Locations in Greater London~\citep{google_maps}. The red dots illustrate the locations of the ATC. The blue triangles illustrate the origin and destination locations specified in order to harvest journey time information.}
	\label{fig:ATC_locations}
\end{figure}

There are 37 DfT ATC locations distributed in Greater London (\cref{fig:ATC_locations}). Among these data collecting locations, 34 roads have ATCs operate in both directions, while three roads have ATCs operate only in one direction. The ATC locations provide traffic counting information for a range of different DfT defined road classes (\cref{tab:ATC_road_class}). The DfT~\citep{DfT_road_classification} publishes guidance on the road classification system in the UK.

\begin{table}[!ht]
	\centering
	\caption{ATC locations by road class}
	\label{tab:ATC_road_class}
	\begin{tabular}{ccccccc}
		\toprule
		Road class & Trunk (Motorway or A) & Principal (A) & B & C & Unclassified & Total \\
		\midrule
		Count & 6 & 16 & 3 & 3 & 9 & 37 \\
		\bottomrule
	\end{tabular}
\end{table}

The raw ATC data contains individual records for each vehicle that passed the counter, including its speed and the exact time (accurate to the second). Over the test period of three weeks, there were approximately 4.5 million recorded vehicles. An example record of a vehicle crossing the ATC Site 11 (labelled in \cref{fig:ATC_locations}) on the first day of data collection reads:

\vspace{2mm}
\textit{Site: 11, {} {}  Direction: Northbound, {} {} Date: 2016-02-27, {} {} Time: 00:00:28, {} {} Speed: 37.}\vspace{2mm}

\noindent
Individual vehicle records at each ATCs are aggregated by the hour to obtain the traffic volumes per hour (i.e., the hourly traffic flow). The total number of vehicles passing in an hour is considered as the hourly traffic volume at the measurement site. For gathering the aggregate data at each ATC, a unique identifier is defined by concatenating the location ID and the directionality of the ATC. For example, the northbound detector at ATC location 11 is identified as site ``11N". The timestamp is then rounded up to the next hour in order to quantify the hourly traffic volume up to that hour. This results in an output dataset which features traffic counts per hour for each site, along with the direction and date. A sample of the aggregated dataset at ATC location 11N is shown in \cref{tab:ATC_agg_data}.

\begin{table}[!ht]
	\centering
	\caption{Processed ATC data record sample}
	\label{tab:ATC_agg_data}
	\begin{tabular}{cccccc}
		\toprule
		Combined ID & ATC ID & Direction & Date & Hour & Hourly Traffic Volume\\
		\midrule
		11N & 11 & Northbound & 2016-03-07 & 6 & 152\\
		11N & 11 & Northbound & 2016-03-07 & 7 & 420\\
		11N & 11 & Northbound & 2016-03-07 & 8 & 694\\
		11N & 11 & Northbound & 2016-03-07 & 9 & 496\\
		\bottomrule
	\end{tabular}
\end{table}

\begin{figure}[htbp!] 
	\centering
	\includegraphics[width=1.0\textwidth]{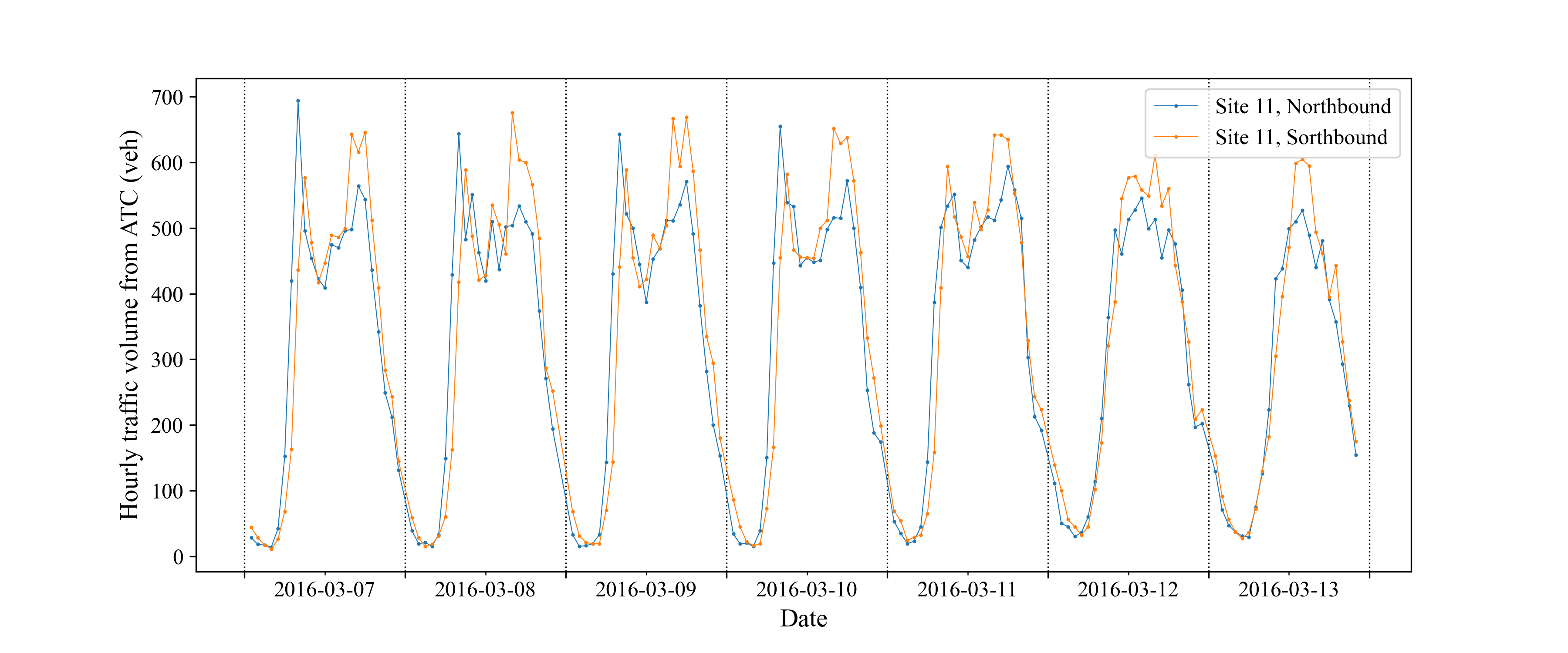}
	\caption{Hourly traffic volume distribution for Site 11 (March 7th-13th, 2016)}
	\label{fig:vehicle_counts_distribution}
\end{figure}

\cref{fig:vehicle_counts_distribution} shows an illustration of the variations in the hourly traffic volume at ATC location 11 (northbound and southbound). Location 11 is situated on the Royal Parade road (A208) with the northbound direction leading to central London and the southbound direction leaving from central London. It is clear that during weekdays (March 7th - 11th, 2016), the northbound direction exhibits a higher peak during the morning rush hours (commuting trips into the city), while the southbound direction carries more traffic leaving the city during the evening peak. On weekends (March 12th and 13th, 2016), there is only one peak, which also starts later than the regular morning peak observed on weekdays. This variation in traffic volume follows the general understanding of the distribution of traffic loads throughout the day.

\section{Traffic speed and time delay from Google Maps API}

Link travel time, or equivalently the inverse of the space-mean speed of vehicles passing through the link, can be collected using several methods. For example, the ATC data presented in the previous section have vehicle speed records. However, the ATC speed is a point measurement and may not be suitable to calculate the average travel time across the road link, as required by many traffic simulation studies. An alternative method to infer the travel time across the link is to use real-time, crowd-sourced location information gathered from mobile phone users. Mobile phones with location service enable harvesting of fine resolution temporal and spatial position data obtained from GPS positioning, cell tower triangulation, WiFi Service Set Identifier (SSID) mapping, Bluetooth, and other technologies, either in isolation or in tandem. Such data hold a great deal of promise due to the range of possible uses it has in the transportation sector, such as understanding peak hour travel demands and inferring transport modes~\citep{ccolak2016understanding,zheng2010understanding}.

\begin{figure}[htbp!] 
	\centering    
	\includegraphics[width=1.0\textwidth]{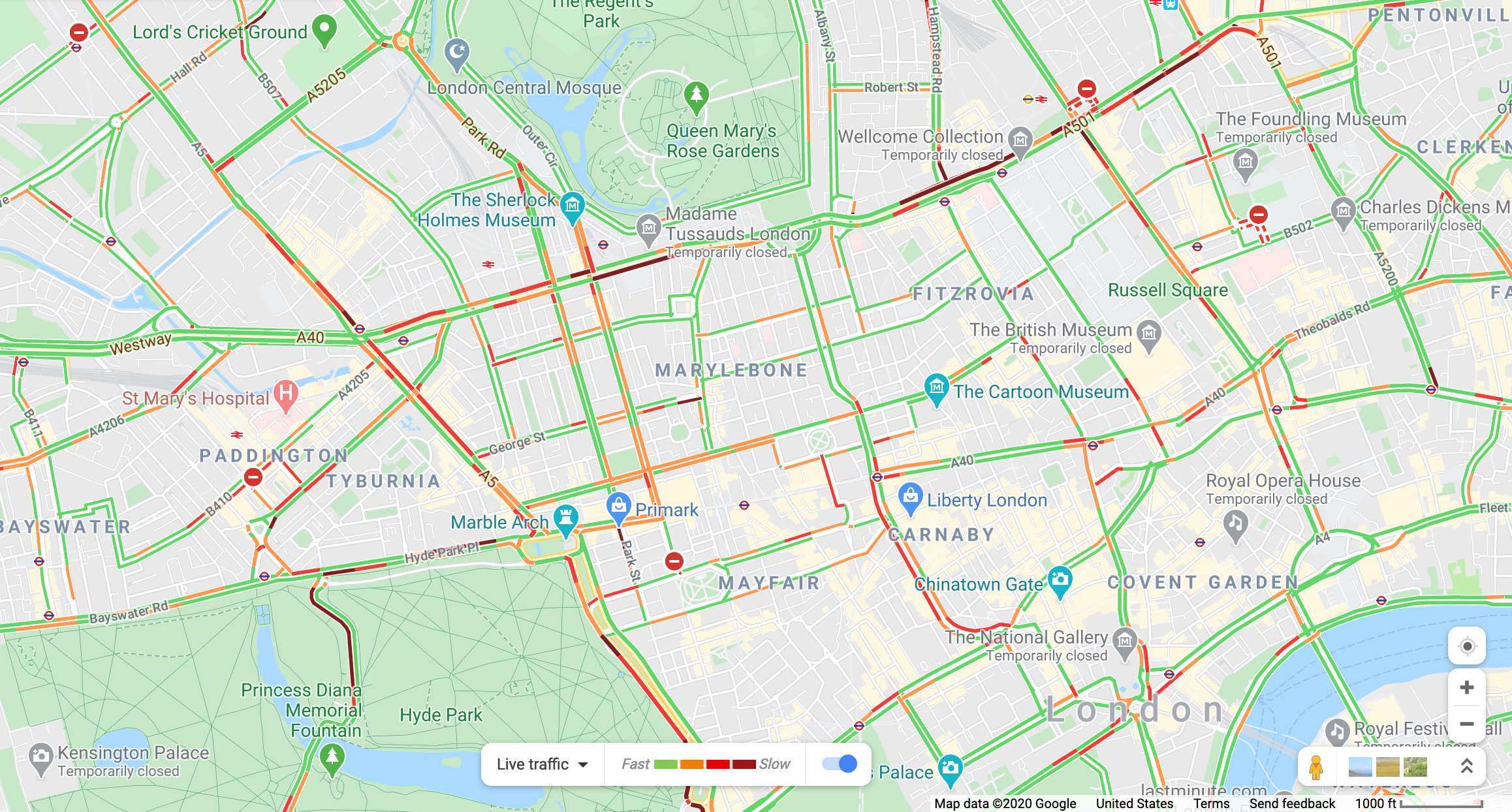}
	\caption[Google Maps Traffic Layer, Camden/Soho/Marylebone/Mayfair area of London ~\citep{google_maps}]{Google Maps Traffic Layer, showing live traffic in the Camden/Soho/Marylebone/Mayfair area of London on a Friday evening~\citep{google_maps}}
	\label{fig:google_traffic_maps}
\end{figure}

Anonymised crowd-sourced location data is useful for understanding real-time road traffic conditions on congested or smooth flowing roads~\citep{google_traffic_blog}. Crowd-source data collection is suitable for urban areas with high population density, high travel demand, and high mobile phone uptake. This information has been widely used by technology companies such as Google and Apple to inform their users of the optimum path that avoids traffic~\citep{apple_maps, bing_maps, google_maps, tomtom_maps}. Individuals with access to these services can make an informed decision on route choice for a given mode or even a mode decision on how to get from their starting point to a desired destination with the lowest time cost. For example, Google provides real-time colour-coded maps that offer a qualitative representation of the current traffic conditions on roads where sufficient data is available. ~\Cref{fig:google_traffic_maps} shows the traffic conditions from Google Maps traffic layer of the Camden area in London on a typical Friday evening.

In this research, the traffic condition information is retrieved in batches using the Google Maps Directions Application Programming Interface (API)~\citep{Google_directions_api}. Depending on the personal settings of a mobile phone user, the freely available Google Maps app send anonymised data of their location to Google. Such data are personally and commercially sensitive, so post-processing is carried out by Google in order to ensure that no user movements can be isolated from the flows. Google's Directions API used in this paper is a service that calculates travel time and routing directions between given origins and destinations using a Hypertext Transfer Protocol (HTTP) request~\citep{Google_directions_api}. The use of an HTTP request allows for scheduled and bulk harvesting of journey information between a selected set of origin and destination pairs. 

This research aims to combine crowd-sourced location-informed journey times with traffic counts from the DfT ATC network. The first step is to specify origins and destinations for the HTTP request to the Google Directions API to retrieve the journey times at the ATC locations. However, the DfT ATC network uses EPSG:27700 (British National Grid) coordinate system, while Google Maps employ the EPSG:4326 (WGS84) system. Hence, there is a need to convert between the two coordinate systems. Furthermore, to retrieve the journey time on a particular road, we need to optimally choose an origin and destination pair whose route would lie on the chosen ATC location section. The choice of origin and destination must be sufficiently further apart such that the journey times are meaningful while having a sufficiently short distance to exclude undesired results such as detours.

\begin{figure}[htbp!] 
	\centering    
	\includegraphics[trim={0 5cm 0 5cm},clip, width=1.0\textwidth]{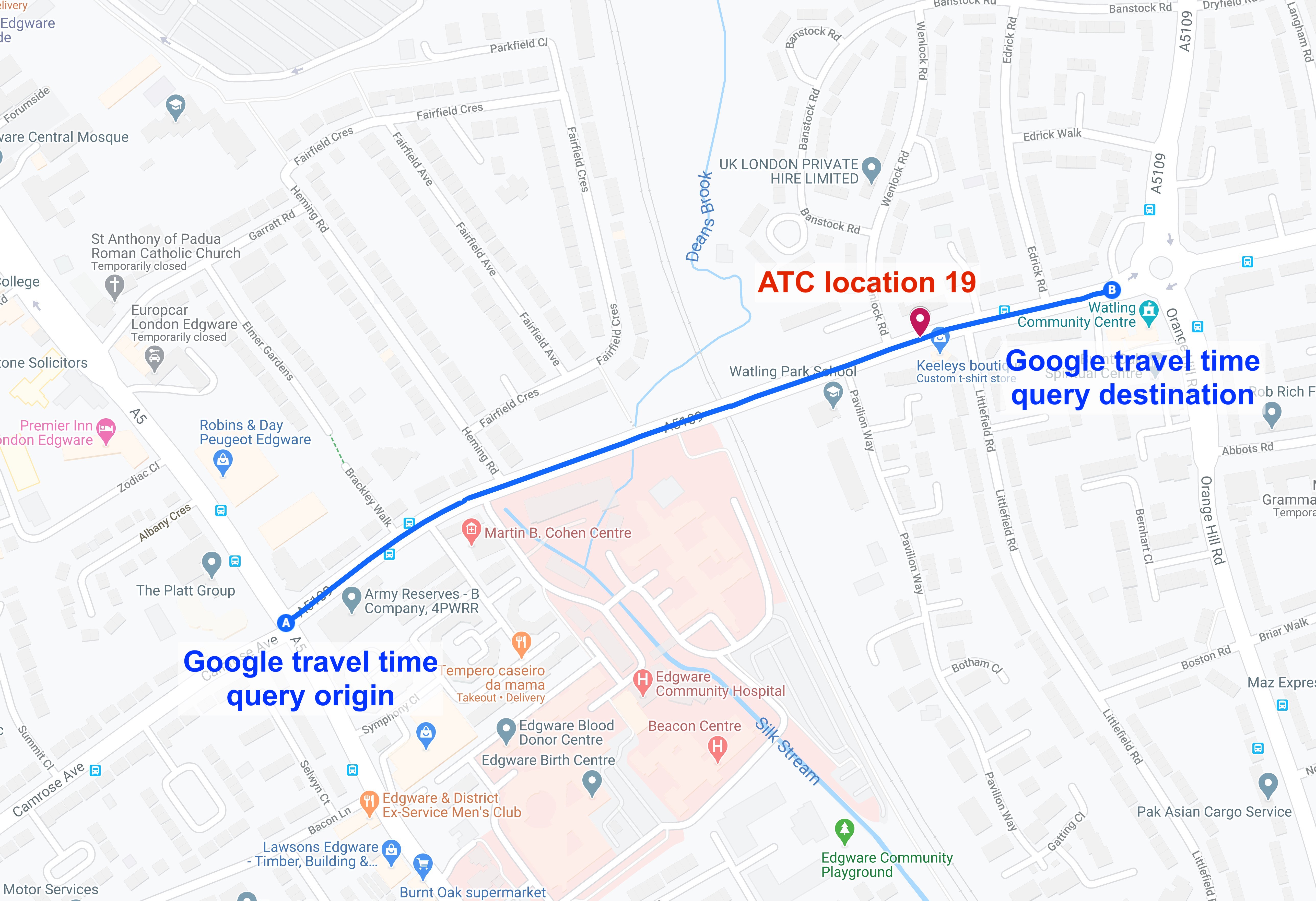}
	\caption{ATC 6 Eastbound with defined origin and destination points~\citep{google_maps}}
	\label{fig:google_od_example}
\end{figure}

\Cref{fig:google_od_example} shows the manually defined origin and destination points for ATC location 19 Eastbound, the A5109 Deansbrook Road in Edgware, HA8. Automating the definition of these origin and destination pairs is challenging as merely taking the start/end of a given road yielded a route that is too long and distorted by other traffic, outside of the ATC consideration. Other methods based upon an idealised distance between points and the density of junctions is deemed too complicated and not durable. For bidirectional ATC locations, it is often not possible to define the Eastbound route as the inverse of the Westbound route as Google distinguish between different sides of the road, resulting in a route which involves a detour to navigate to the correct orientation safely. Thus a manual process is employed to visually inspect each location, the surrounding context and decide on the most appropriate origin and destination locations for querying for the travel time. Once this manual process is complete, a list of origin and destination pairs is produced containing ATC metadata that allows for the pairing of Google travel time results to its corresponding ATC traffic volume data. 

During the information request process, an HTTP request is sent to Google's Directions API with the following information: origin, destination, mode (driving) and the specified departure time. In order to harvest real-time data that is informed by aggregated data of individual device-location at each hour, a cron scheduler~\citep{cron} is used to run the same origin and destination pairs repeatedly. In response to such HTTP request,  Directions API returns results in the JavaScript Object Notation (JSON) format, a lightweight data-interchange format~\citep{json}. The JSON result has the $duration\_in\_traffic$ field, which is Google's estimated journey time between the requested origin and destination. In the next section, we explain how these travel time data will be paired with ATC vehicle counts to produce volume-delay relationships.

\begin{figure}[!htbp] 
	\centering
	\includegraphics[width=1.0\textwidth]{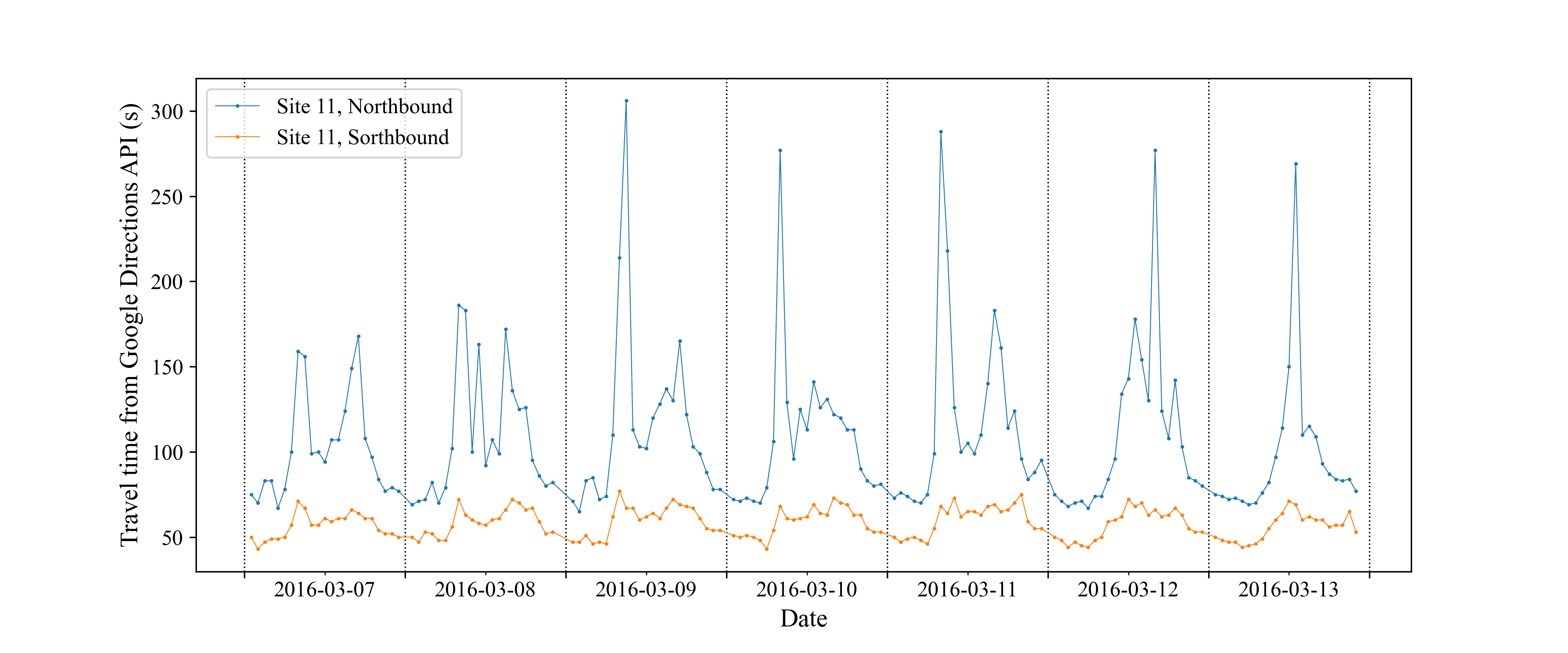}
	\caption{Journey time distribution for Site 67 (March 7th-13th, 2016) along the Royal Parade, between Manor Park Road and Bromley Road in Southeast London (561 metres in length)}
	\label{fig:travel_time_distribution}
\end{figure}

\cref{fig:travel_time_distribution} shows a subset of Google's estimated journey time data along the Royal Parade, between Manor Park Road and Bromley Road in Southeast London (561 metres in length). ATC 11 is located on the north end of this stretch of the road. Apart from the bimodal distribution of journey time with peaks associated with the morning (to work) and evening (leaving work) periods during weekdays and the single peak on weekends~\citep{mullick2012dynamics}, it is interesting to see a much higher travel time in the northbound direction at all times. A visual inspection of the corresponding road shows the presence of on-street parking spots in the northbound lane. Also, the ending point of the northbound lane intersects with a more major road (Bromley Road, A222) than southbound (Manor Park Road, B264). Both factors could contribute to the unexpected slowness in the northbound direction.

\section{Data cleaning and site selection} \label{sec:cleaning}
\begin{figure}[htbp!] 
	\centering
	\includegraphics[width=\textwidth]{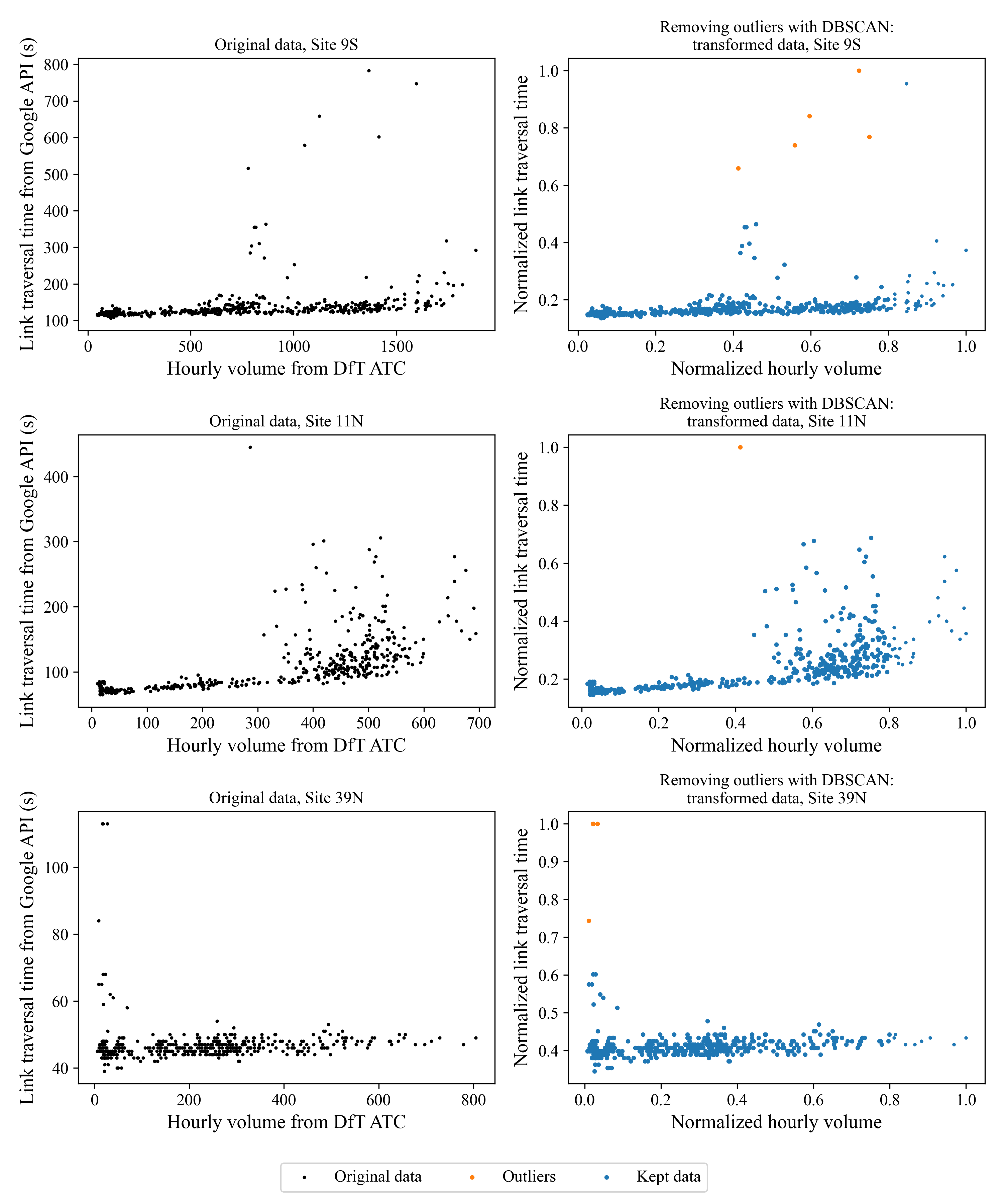}
	\caption{Examples showing the original, removed and kept observations in volume-delay scatter plot for three sites.}
	\label{fig:clean}
\end{figure}

The volume delay relationship at different traffic conditions at each site is constructed by combining the Google travel time data and the ATC vehicle count data. \cref{fig:clean} shows an example of such data for site 9S, 11N and 39N. Outliers with high travel times can be observed at moderate traffic flow rates. These abnormal delays could occur due to external reasons (e.g., weather and road incidents), or as mentioned in the literature review, being the hyper-congested cases where flow reduces with increase in delay (especially the cluster of points with hourly volume between 400 and 600 for site 11N). The hyper-congested regime is not modelled by the volume-delay function. Instead, they are treated as residuals of the fitted volume-delay curve and quantifications of the magnitude of such residuals are provided in~\cref{sec:mae}. 

Regardless of the residual quantification, the collected data must be cleaned to remove points falling too far away from the continuous volume-delay relationship. Specifically, a manually tuned Density-based spatial clustering of applications with noise (DBSCAN) algorithm is used to remove points in low-density areas in the plots~\citep{Brownlee2020, scikit2019}. The DBSCAN algorithm has two parameters, $\epsilon$, which sets the distance between points to be considered as a cluster and \textit{minPoints}, the minimum number of points required to be considered as a core point in the cluster. Considering that the horizontal and vertical axes have different scales, the original data are first normalized to the maximum observed volume and travel time, so that the data in the transformed axes are all within 0 and 1. It was found that by specifying $\epsilon=0.1$ and $minPoints=5$, it offers a good balance between removing outliers and keeping data points close to the main cluster. In fact, data cleaning is purposefully kept to a minimum to accommodate possible variations that may occur in real life. As a result, only 66 points are removed from the 17,745 observations.

Furthermore, when the journey time versus traffic counts data for all the 39 original sites were plotted, it was immediately obvious that a few sites do not have valid data (very low or no variations in observed travel time through the observation period). Most of these sites only have very low traffic volume throughout the data collection period. For example, for site ``74S", Swain's Lane, a minor road in Camden, the maximum hourly volume is only 24 vehicles. The travel delays on these roads are relatively small and the data collected cannot be used to extract knowledge on the delays that might occur when the road is congested. These sites are subsequently removed from analysis and which leaves 24 sites remaining for further analysis.

\section{Analysis and model building}

The combination of Google travel time data and the DfT ATC vehicle count data enables various relationships to be assessed and defined. Based on these real-world datasets, the goal of this section is to derive data-driven parameters related to the traffic analysis, including (1) the free-flow travel time, or its inverse, the free-flow speed; (2) the road capacity; and (3) the volume-delay relationship.

\subsection{free-flow travel time}\label{sec:speed}
\begin{figure}[htbp!] 
	\centering    
	\includegraphics[width=1.0\textwidth]{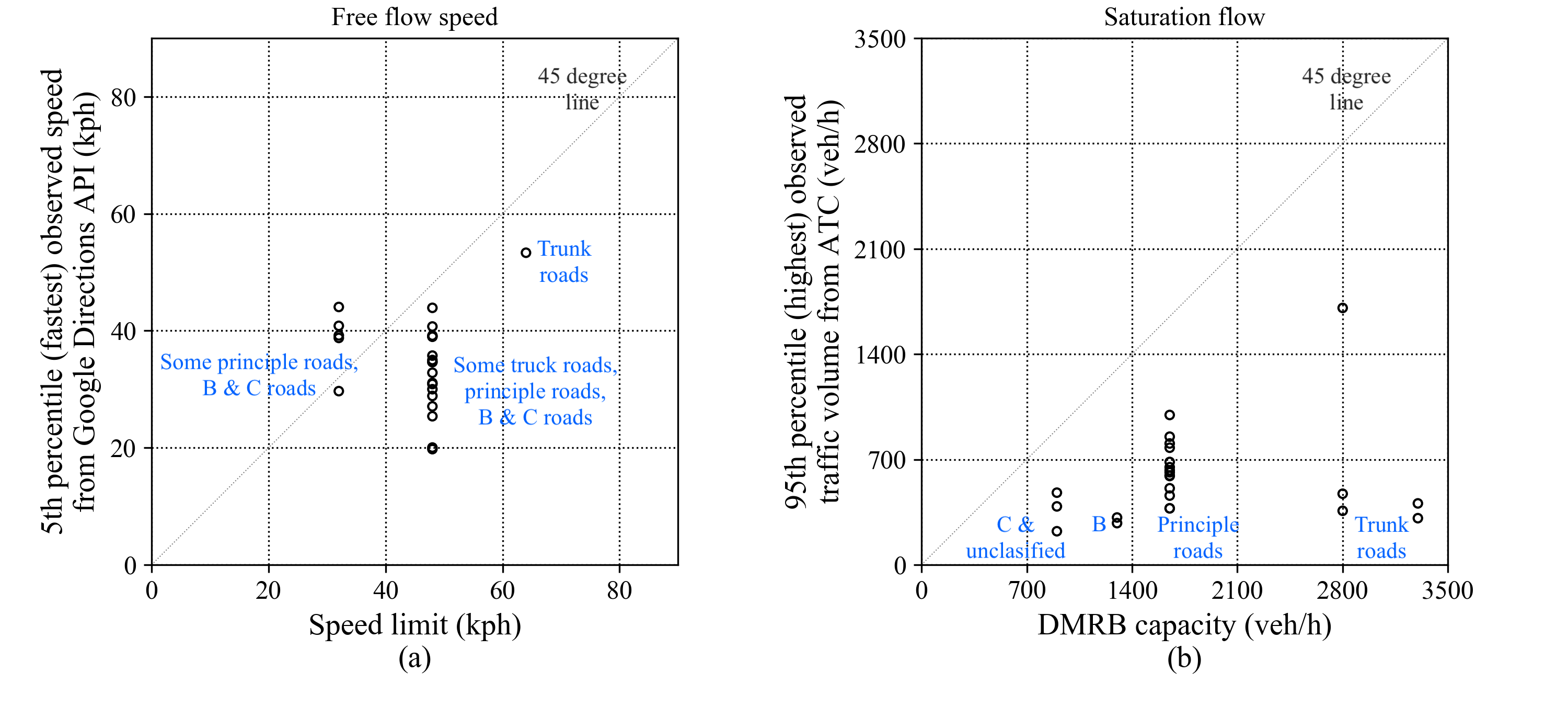}
	\caption{Comparison between the theoretical traffic parameter value and observed value at 24 ATC sites. (a) Free-flow speed; (b) capacity, or saturation flow.}
	\label{fig:fft_cap_design_observe}
\end{figure}

The free-flow travel time usually corresponds to the time for a vehicle to pass through a road link when no other vehicles are present, such as the travel time experienced in the early morning. In research studies and engineering practice, it is usually taken as the time to go through a link when travelling at the designated speed limit, sometimes multiplied by a slowing-down factor of roughly 1.3 to reflect the minor delays caused by stopping at intersections and other factors in an urban environment~\citep{ccolak2016understanding}. For example, the DfT defines the free-flow journey time using time delay coefficients for different road types, speed limits, link lengths, widths, gradients and traffic junctions, which is obtained for our study sites and plotted as the horizontal axis value in \cref{fig:fft_cap_design_observe} (a)~\citep{DfT_cost_benefit}.

In this study, it is recognised that using speed limit to calculate the free-flow travel time based on road type frequently ignores the impact of localised, irregular factors, for instance, a curve or pothole on the road. Rather than making assumptions about the free-flow time, we estimated it from the Google travel time data collected over the period of the study. For each site, the 5th percentile (fastest) travel time among all observation points is used to represent the observed free-flow travel time of that particular site. Free-flow speeds for each direction of the same road are estimated separately. \cref{fig:fft_cap_design_observe}(a) shows the contrast between the speed limit and the observed maximum speed calculated from the observed free-flow travel time. Each point represents a specific site in the dataset. On average, the observed maximum speed of a road link is 20\% lower than the specified speed limit. However, the former tends to be higher (exceeding the speed limit) on roads with low posted speed limit. A 45-degree line is also plotted in~\cref{fig:fft_cap_design_observe}(a) for reference to compare the speed limit against the observed maximum speed.

\subsection{Road capacity}\label{sec:capacity}

Compared to the free-flow travel time, road link capacity is an ambiguous quantity to define. For example, the UK Design Manual for Roads and Bridges (DMRB) defines capacity vaguely as ``the maximum sustainable flow of traffic passing in one hour, under favourable road and traffic conditions"~\citep{DMRB}. The DMRB also provides lookup tables that feature traffic capacities for a range of road types, road widths and number of lanes. A manual survey of satellite imagery was carried out to assess the lane count, estimate the road width for each of the ATC locations and thus provide an estimated capacity by this DMRB definition. However, such a method was deemed unacceptable due to the uncertainty in what constitutes as \textit{favourable road and traffic conditions}.

Instead, a more nuanced definition from Spiess~\citep{spiess1990conical} is adopted, which defines capacity as the volume at which congested speed is half the free-flow speed. In the combined Google and ATC data, since there is hardly any observation when the speed is exactly half of the free-flow speed, a more relaxed definition of capacity is subsequently utilised. For each site, the capacity is taken as the highest 95\textsuperscript{th} percentile hourly traffic volume among all observations where the travel time falls between 1.8 to 2.2 times the free-flow speed. Capacities for each direction of the same road are estimated independently, since each direction may have different contexts, such as the presence of bus stop, street parking or lane width. As a result, each individual road is given a capacity attribution. This capacity attribution is observed to be consistently lower than the DMRB recommended value, as can be seen in~\cref{fig:fft_cap_design_observe} (b). On average, the observed capacities are only 34\% of those recommended by the DMRB, reflecting factors such as signalised intersections, bus stops and street parking that cannot be perfectly captured using simple lookup tables. The 45-degree line is plotted in~\cref{fig:fft_cap_design_observe} (b) for reference. Also, it can be seen that there are only a few values of the DMRB road capacity, clustered according to the road types and lane counts.

\subsection{Volume-delay relationship}
Based on observations of the road link travel time and hourly traffic counts presented above, the context-specific volume-delay relationship can be constructed. The BPR formulation of the volume-delay relationship is chosen due to its mathematical advantages (differentiable and convex) and widespread use (\cref{eq:BPR}). In particular, for the baseline case, we used $\alpha=1.0$ and $\beta=2.0$ as per TfL recommendation, while $t_0$ and $v_c$ are based on the speed limit and DMRB recommended capacity as computed in \cref{sec:speed} and \cref{sec:capacity}. It should be noted that when the BPR curve is used in practice, $t_0$ is often corrected by a delay factor to reflect the city-specific conditions~\citep{ccolak2016understanding} while $v_c$ is often corrected by factors such as road geometry, gradient and signal status~\citep{kurth1996implementation}. However, these corrections usually require labour-intensive site inspections and the correction factors are often empirical or generalised for the whole city. The aim of this analysis is to propose an efficient way to obtain alternative, site-specific function parameters from real observations. The purpose of this study is to also contrast the performance of the data-informed model parameters with the base curve in terms of their ability in calculating the time-delay at different congestion levels.

A summary of the three models used for comparison is given in \cref{tab:functions}. The base curve uses BPR coefficient $\alpha=1$ and $\beta=2$. In particular, the free-flow travel time $t_0$ is calculated from the speed limit, while the capacity $v_c$ follows the DMRB recommended value. In traffic modelling, the transport agencies may apply some adjustment coefficients to consider external influencing factors on $t_0$ and $v_c$. However, in most of the cases, these factors are predetermined and may not be up-to-date or localised to each road segment. The alternative formulations are developed to address this issue: for the first data-informed alternative formulation (DD1), the base value of $t_0$ and $v_c$ are replaced by observed values obtained in \cref{sec:speed} and \cref{sec:capacity}. For the second data-informed alternative (DD2), not only $t_0$ and $v_c$ are replaced by the observed values, but also $\alpha$ and $\beta$ are calibrated using regression models on the real-world data. Once the data collection and processing pipeline have been built, it is relatively easy to obtain the coefficients used in the data-informed volume-delay curves, even updating weekly or monthly to take into account any temporal changes in the road characteristics (e.g., roadworks or weather conditions).

\begin{table}[!ht]
	\centering
	\caption{Summary of volume-delay relationships tested.}
	\label{tab:functions}
	\begin{tabular}{p{4cm} p{8cm}}
		\toprule
		Function name & Parameters \\
		\midrule
		Base curve & $\alpha=1.0$, $\beta=2.0$ \\
		             & $t_0$: road length divided by speed limit. \\
		             & $v_c$: DMRB definition. \\
		             & \\
		Data-informed 1 (DD1) & $\alpha=1.0$, $\beta=2.0$ \\
		                & $t_0$: 5th percentile link travel time (fastest) in observations. \\
		                & $v_c$: 95\textsuperscript{th} percentile traffic volume (largest) for observations with travel time between 1.8 to 2.2 $\times t_0$. \\
		                & \\
		Data-informed 2 (DD2) & $\alpha$, $\beta$: fitted value using non-linear least square regression. \\
		                & $t_0$: 5th percentile link travel time (fastest) in observations. \\
		                & $v_c$: 95\textsuperscript{th} percentile traffic volume (largest) for observations with travel time between 1.8 to 2.2 $\times t_0$. \\
		\bottomrule
	\end{tabular}
\end{table}

It is evident from \cref{fig:clean} that the observed relationships between the volume and delay are site-specific, non-linear and heteroskedastic (variance of the data increases with the independent variable). All these characteristics are expected from the traffic perspective. In particular, the non-linearity means that the travel time increases more dramatically as the hourly traffic volume approaches the capacity. The heteroskedasticity implies more variations in vehicle speed when the road is congested due to interactions with other vehicles. Usually, to deal with such data, a non-linear model needs to be used (which is already the case for the BPR curve). Heteroskedasticity can be accounted for by using techniques that include data transformation or using Weighted Least Square (WLS) method~\citep{vynck2017heteroscedasticity}. However, it is not a requirement if the goal is to estimate regression parameters, as the Ordinary Least Square (OLS) method also produces unbiased, though inefficient, estimation of the coefficients. Standard errors of the estimation parameters are not used, so the OLS method is deemed sufficient for obtaining the $\alpha$ and $\beta$ parameters used in the DD2.

\begin{figure}[htbp!] 
	\centering    
	\includegraphics[trim={0 0cm 0 1cm},clip, width=1.0\textwidth]{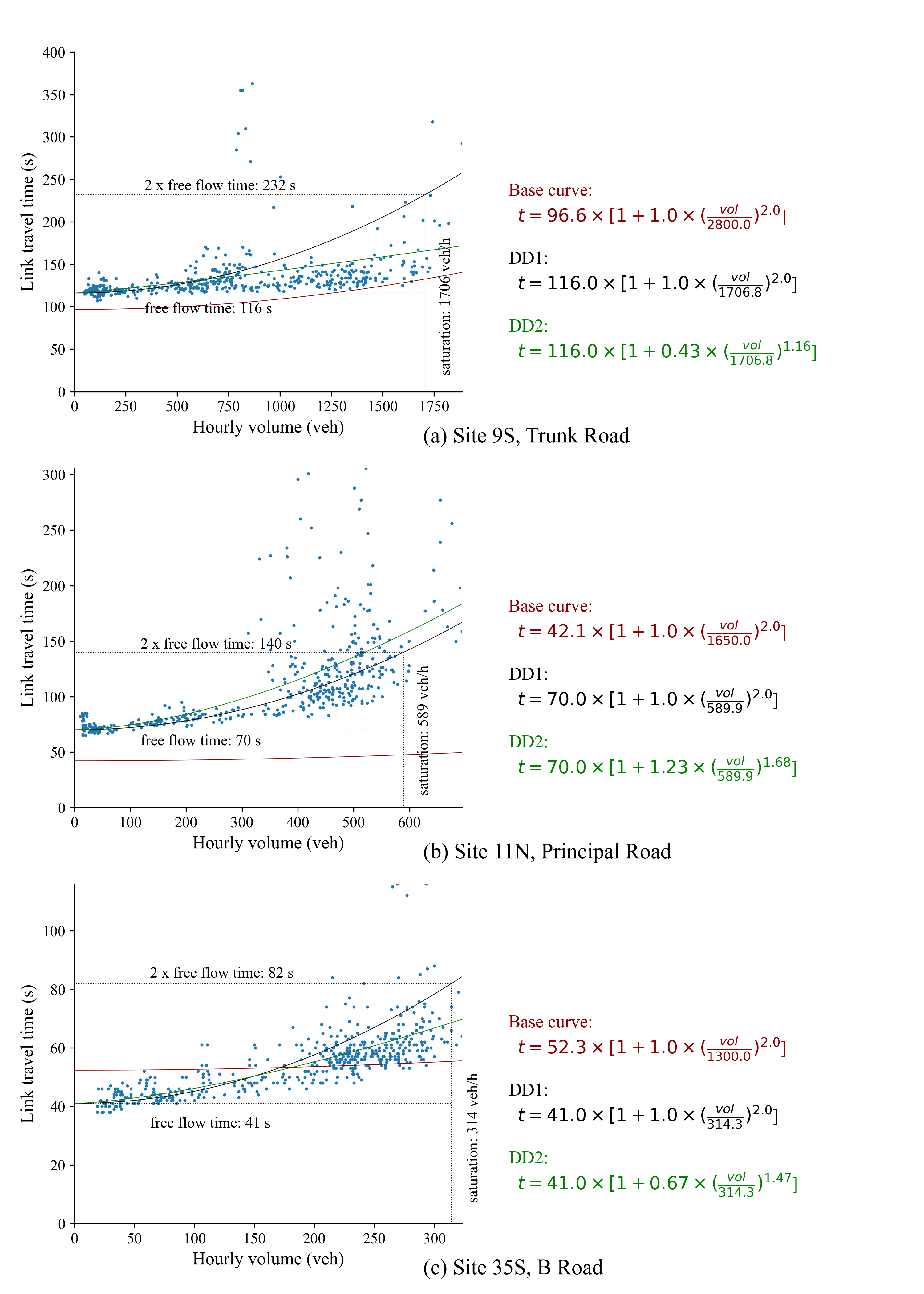}
	\caption{The fitting three different models in the observed data at three sites.}
	\label{fig:curve_fit}
\end{figure}

\cref{fig:curve_fit} shows the fitting results of all candidate functions for three representative sites, each from a different road class. It can be seen that the base curves with the default speed and capacity values fit the poorest to the data across all three examples. In particular, the base curves underestimate the free-flow time for Site 9S (\cref{fig:curve_fit} (a), Trunk road) and 11 N (\cref{fig:curve_fit} (b), Principal road), while overestimate the free-flow time for Site 35S (\cref{fig:curve_fit} (c), B road). The base curves overstate the capacity in all three examples, which has already been shown in \cref{fig:fft_cap_design_observe} (b). By simply substituting the free-flow time $t_0$ and capacity $v_c$ in the base curve with the observed values, the fitting of the DD1 improves significantly based on a visual inspection. Quantitative evaluations of the fitting performance will be given next in \cref{sec:mae}. The DD2 (green curve in \cref{fig:curve_fit}) fits the data more closely than the DD1. However, in certain cases, the DD2 may result in unrealistic $\beta$ values of less than one, due to a large number of observations at medium volume but large delay such as the hyper-congestive conditions, thereby making the volume-delay curve unreasonably concave. Hence, the DD1 is a good approach in general to improve the performance of volume-delay curves.

\begin{figure}[htbp!] 
	\centering    
	\includegraphics[trim={0.5cm 0cm 0cm 0cm},clip, width=1.0\textwidth]{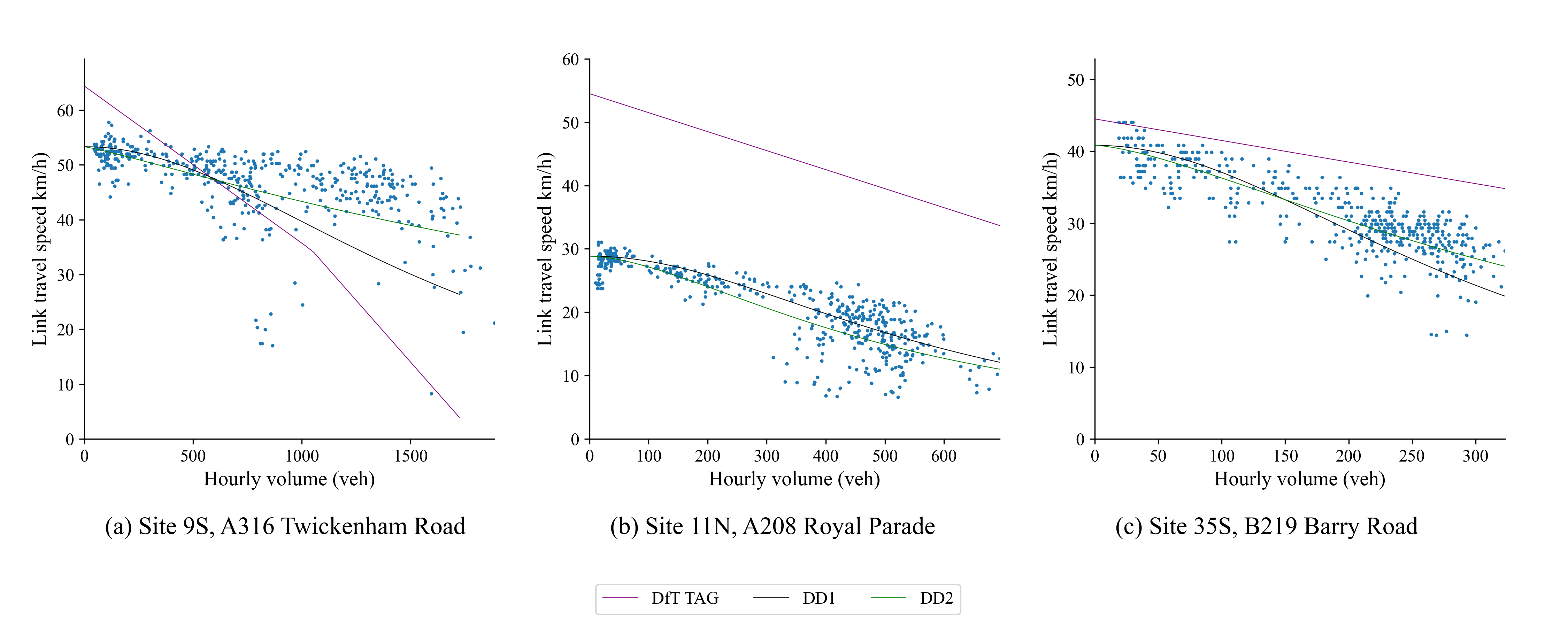}
	\caption{Comparison with the DfT TAG speed-flow curves for three sites.}
	\label{fig:curve_fit_speed_horizontal}
\end{figure}

By dividing the length of the road link with the link travel time shown in \cref{fig:curve_fit}, the speed-flow relationship can be obtained. The DfT TAG uses linear or piece-wise linear models to predict the reduction in speed with increasing flow. Depending on the road class (e.g., whether it is urban, suburban, rural), different formulae are used. For each class, the speed-flow formulae again depend on a variety of factors (e.g., percentage of frontage development, numbers of major and minor intersections). It should be noted that, although a speed-flow formula is provided for urban roads, the preferred way of doing traffic assignment is still to model junctions and low cruise speed in congested urban areas explicitly, rather than relying on the speed-flow curves. If to be used, it should be recognized that the speed-flow curves are developed based on network-average conditions, which limits their applicability to individual links independently. Fully recognizing these limitations, a comparison plot is still made for the observed data points and the speed-flow curves in the TAG document for illustration purpose (\cref{fig:curve_fit_speed_horizontal}). The blue dots are the speed-flow observation data. The black and green curves are the same as the ones in \cref{fig:curve_fit}, only showing in the speed-flow plane. The purple lines are the speed-flow relationship made according to the TAG model. Parameters of the purple curves are given in \cref{tab:functions_dft_tag} and these values are obtained by examining Google Street View of each site presented in \cref{fig:curve_fit_speed_horizontal}. It can be seen that, the purple curves do not agree well with the local data observations. Part of it is because link observations (e.g., the percentage of frontage development for this particular street) are used in place of network-average values. While it is also true that the curve parameters, even if to be collected at the network level, are somewhat subjective/elastic and may not represent all context-specific information (e.g., bus stops, subtle geometry effects, or temporary closures). In comparison, the proposed crowd-sourced data collection approach can then provide a more localised, context-specific estimation of the volume-delay curves, or even used to calibrate parameter values of the TAG models.

\begin{landscape}
\begin{table}[!ht]
	\centering
	\small
	\caption{Parameter values for curves in \cref{fig:curve_fit_speed_horizontal}}
	\label{tab:functions_dft_tag}
	\begin{tabular}{p{2cm} p{5cm} p{5cm} p{3cm} p{3cm}}
		\toprule
		Curve & Basic form & Site 9S & Site 11N & Site 35S \\
		\midrule
		DfT TAG \newline (purple curve) 
		    & $v = v_0 - coefficient \times volume$
		    & Class 10,\newline suburban single carriageway 
		    & Class 7,\newline urban non-central roads 
		    & Class 7,\newline urban non-central roads \\
		    && Percentage of heavy vehicles\newline $PHV(\%)=0$ 
		    & Frontage development\newline  $DEVEL(\%)=50$ 
		    & Frontage development\newline  $DEVEL(\%)=100$ \\
		             && Maximum realistic flow\newline $Q_C=1500*(92-PHV/80)$ & Free-flow speed\newline $v_0=64.5-\frac{DEVEL}{5}$ & Free-flow speed\newline $v_0=64.5-\frac{DEVEL}{5}$  \\
		             && Flow change point\newline $Q_B=0.7 \times Q_C$ & $v=v_0-30\times\frac{volume}{1000}$ & $v=v_0-30\times\frac{volume}{1000}$  \\
		             && Major intersections\newline $INT=1/km$ & & \\
		             && Minor intersections\newline $AXS=4/km$ & & \\
		             && Free-flow speed\newline $v_0=70 - 5 \times INT - 3 \times AXS/20$ & & \\
		             && For flow $<Q_B$\newline $v=v_0 - (12 + 50 \times \frac{INT}{3}) \times \frac{volume}{1000}$ & & \\
		             && For $Q_B<$ flow $<Q_C$\newline $v=v_0 - (12 + 50 \times \frac{INT}{3}) \times \frac{Q_B}{1000} - 45 \times \frac{volume-Q_B}{1000} $ & & \\
		             && & & \\
		             && & & \\
		DD1 \newline (black curve)
		    & $v=v_0 / (1+\alpha \times (\frac{volume}{capacity})^{\beta})$
		    & $capacity=1706.8$\newline$v_0=53.3$ 
		    & $capacity=589.9$\newline$v_0=28.9$ 
		    & $capacity=314.3$\newline$v_0=40.8$ \\
		        && $\alpha=1$, $\beta=2$ & $\alpha=1$, $\beta=2$ & $\alpha=1$, $\beta=2$\\
		             & & & & \\
		DD2 \newline (green curve)
		    & $v=v_0 / (1+\alpha \times (\frac{volume}{capacity})^{\beta})$
		    & $capacity=1706.8$\newline $v_0=53.3$ & $capacity=589.9$\newline $v_0=28.9$ & $capacity=314.3$\newline $v_0=40.8$ \\
		        && $\alpha=0.43$, $\beta=1.16$ & $\alpha=1.23$, $\beta=1.68$ & $\alpha=0.67$, $\beta=1.47$ \\
		\bottomrule
	\end{tabular}
\end{table}
\end{landscape}

\subsection{Variation quantification}\label{sec:mae}
As it can be seen in \cref{fig:curve_fit}, the real-world traffic data are very noisy to be represented by a single curve. There have been second-order models that include a family of curves, which can reproduce the variations in travel time around the same traffic flow conditions to a certain degree~\citep{generic2020fan}. However, well-calibrated first-order models that only include the speed term (no time derivatives of it), such as the BPR curve, are still beneficial for specific tasks, including analysing the impact of building new infrastructures and conducting regional-wide static and semi-static traffic simulations. After fitting the three candidate models to the data, their fitting performances are analysed in this section.

\begin{figure}[htbp!] 
	\centering    
	\includegraphics[trim={1cm 1cm 3cm 2cm},clip,width=1.0\textwidth]{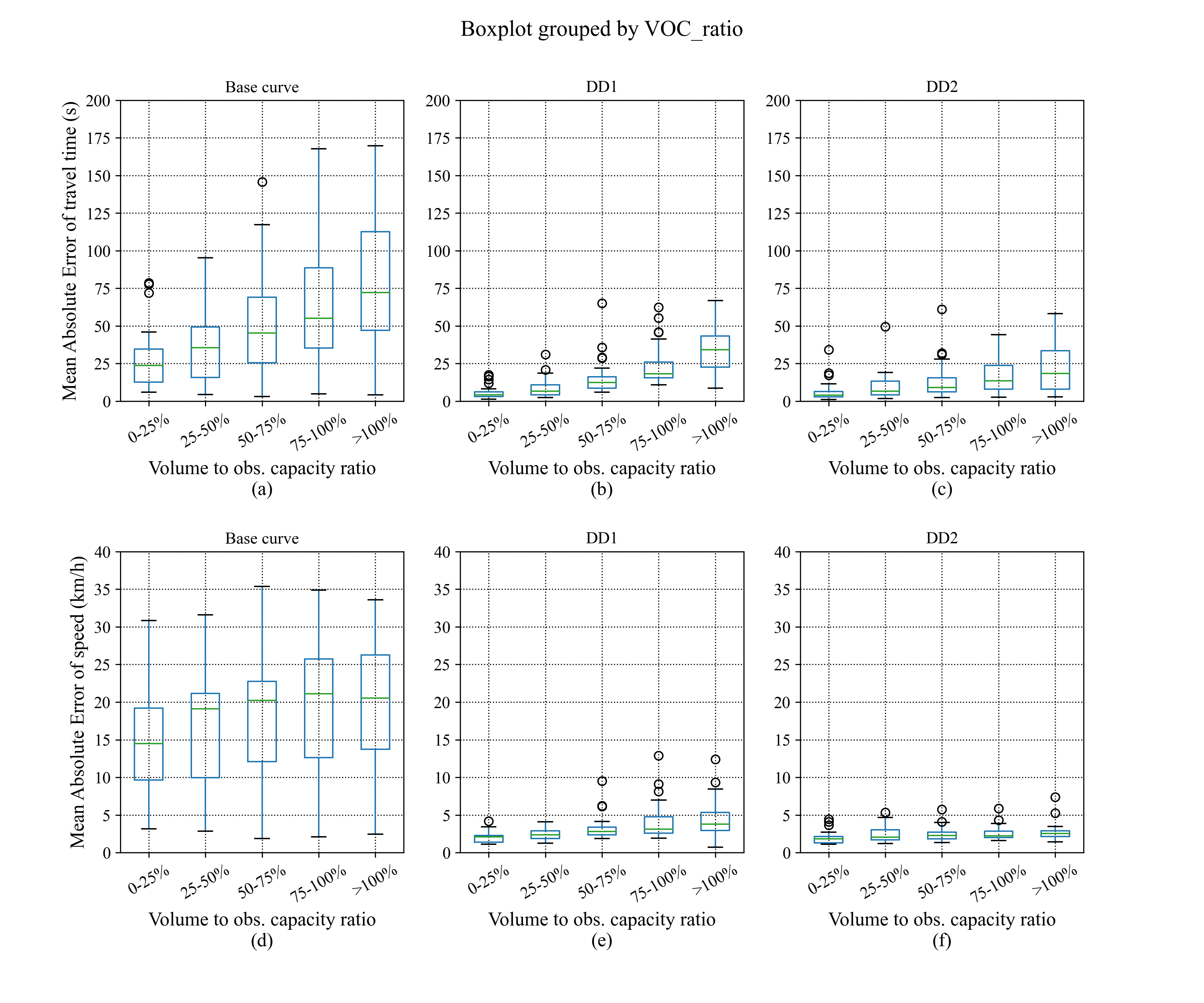}
	\caption{Boxplots of the overall fitting MAE for all study sites for three model types, grouped by the traffic volume level. (a-c) MAE for the estimated travel time. (d-f) MAE for the estimated link speed.}
	\label{fig:mae}
\end{figure}

The fitting performance is evaluated with the Mean Absolute Error (MAE, \cref{eq:MAE_eq}) because the MAE metric has the same unit as the dependent variable (time in seconds or speed in km/h) and is less sensitive to outliers compared to the frequently used Root Mean Squared Error (RMSE) metric. As the variations in the observed travel time are uneven and more substantial when the hourly traffic volume approaches the capacity, MAE is calculated for different subsets of the data, divided based on the quartile of the hourly traffic volume. \cref{fig:mae} shows the MAE in terms of travel time (a)-(c) and speed (d)-(f) estimations of all roads, grouped by the volume-to-capacity ratio of the observed data. For example, Q1 refers to the subset of the data where the hourly traffic volume are within 0-25\% of the estimated capacity. Q5 refers to the subset of the points where the hourly traffic volume exceeds the estimated saturation flow. The Q5 partition exists because capacity in this study is defined as the 95\textsuperscript{th} percentile hourly traffic volume for observations with travel times between 1.8 to 2.2 times the free-flow time~(\cref{sec:capacity}). It can be seen that, in terms of travel time or speed, the base curve leads to the largest errors in estimating the time delay given the traffic volume (\cref{fig:mae} (a) and (d)). While for the two data-informed models, the errors of the DD2 appear to be less, but not much lower than the DD1. As mentioned before, engineers and planners would normally apply a series of factors on $t_0$ and $v_c$ for the base curve to reflect city or street characteristics (e.g., signal green time ratio or road curvature). We tried to apply these factors to scale $t_0$ and $v_c$ so that the base value match more closely with the observed value in \cref{fig:fft_cap_design_observe} , but it still produces larger MAE compared to the two data-informed curves due to its lack of ability to adapt to individual street situations.

\begin{equation}\label{eq:MAE_eq}
\begin{split}
    \text{For travel time: } & MAE_t = \frac{\sum_{i=1}^{n} |t_{i,model} - t_{i,obs}|}{n} \\
    \text{For speed: } & MAE_u = \frac{\sum_{i=1}^{n} |u_{i,model} - u_{i,obs}|}{n}
\end{split}
\end{equation}
\noindent
where $t_{i,model}$ and $u_{i,model}$ are estimated travel time and speed from the model; $t_{i,obs}$ and $u_{i,obs}$ are individual travel time and speed observations; $n$ is the total number of observations.

The above results not only indicate the performance of the models, but also give engineers quantitative evidence on what amounts of variations to be expected when using the simple volume-delay function curves in their estimation. For example,~\cref{fig:mae}~(a)-(c) shows the variations between the observed link traversal time and the estimated travel time by each individual model. For the roads selected as the study sites, the link travel time predictions can be 1-2 minutes off using the base curve, while the errors are usually below half a minute using the data driven models (DD1 and DD2), except in Q5, where the traffic volume is relatively high (> observed capacity) and the errors increase to 0.5-1 minute. The model prediction errors are better discussed by comparing the predicted speed against the observed speed at different traffic levels, as in~\cref{fig:mae}~(d)-(e), since the speed measure can be compared across roads. As seen in~\cref{fig:mae}~(d)-(e), without the site-specific knowledge, the speed estimations using the default volume-delay curve can vary by 10-25 km/h (\cref{fig:mae} (d)). Nevertheless, merely fixing the speed and capacity estimations (DD1), the variation in the speed estimation can be reduced to less than 5 km/h for low to medium traffic volume cases, and less than 10 km/h when the volume is close to the capacity (\cref{fig:mae} (d)). If using calibrated coefficients $\alpha$ and $\beta$ as in the DD2, the variation in the speed estimations are almost always below 5 km/h from the observations (\cref{fig:mae} (e)). However, it should be noted that calibrating $\alpha$ and $\beta$ without constraints may produce concave volume-delay curves in the case of the DD2.

\cref{tab:mae_summary} shows the MAE of speed (km/h) by model, road class and volume level, which is an extension of the information in \cref{fig:mae} (d)-(f) disaggregated by the road facility type. As can be seen in the table, the Principal roads generate the highest number of observations, while the Unclassified roads have the least. Among the observations from Principal roads, the largest group of observations have the volume-to-capacity ratio around 75-100\%, indicating a high demand. At the same time, less important roads (e.g., C or Unclassified) have the majority of the observations in the low traffic category (less than 75\%), which means less congestion in general on these types of roads. The general trend on the MAE for the estimated speed is not much different across different road classes. For all road classes, the MAE of the estimated speed is significantly higher when using the base curve and lower when using the data-informed models. Also, the MAE for speed, in general, grows with the volume level: when the traffic volume-to-capacity ratio is higher, the MAE for speed goes up as well, which is in agreement with the visual judgements made in \cref{fig:mae}.
\begin{table}[!ht]
	\centering
	\caption{Speed MAE (km/h) by model form, road type and traffic volume level}
	\label{tab:mae_summary}
	\begin{tabular}{cccccccc}
		\toprule
		\multicolumn{2}{c}{Volume-to-capacity ratio} & <25\% & 25-50\% & 50-75\% & 75-100\% & >100\% & Total \\
		\midrule
		\multirow{5}{*}{Trunk} & obs. counts & 383 & 245 & 375 & 366 & 23 & 1392 \\
		& \% obs. counts & 27.51 &	17.60 & 26.94 &	26.29 &	1.65 &	100 \\
		& Base & 14.4 & 19.3 & 23.5 & 25.8 & 27.9 & 19.3 \\
		& Data 1 & 2.3 & 3.3 & 2.7 & 2.4 & 3.8 & 2.7 \\
		& Data 2 & 1.7 & 2.6 & 2.3 & 2.2 & 2.8 & 2.2 \\
		& & & & & & & \\
		
		\multirow{5}{*}{Principal} & obs. counts & 1641 & 1082 & 1996 & 2746 & 216 & 7681 \\
		& \% obs. counts & 21.36 & 14.09 & 25.99 & 35.75 & 2.81 &	100.00 \\
		& Base & 13.3 & 17.2 & 18.9 & 21.1 & 21.5 & 18.7 \\
		& Data 1 & 1.9 & 2.2 & 2.9 & 2.8 & 3.4 & 2.6 \\
		& Data 2 & 1.8 & 2.0 & 2.5 & 2.6 & 2.5 & 2.4 \\
		& & & & & & & \\
		
		\multirow{5}{*}{B} & obs. counts & 179 & 125 & 183 & 343 & 102 & 932 \\
		& \% obs. counts & 19.21 & 13.41 & 19.64 & 36.80 & 10.94 & 100.00 \\
		& Base & 9.0 & 6.6 & 4.3 & 3.5 & 3.8 & 4.8 \\
		& Data 1 & 1.8 & 2.3 & 4.6 & 6.7 & 8.6 & 3.6 \\
		& Data 2 & 1.8 & 2.1 & 2.2 & 2.1 & 2.2 & 2.1 \\
		& & & & & & & \\
		
		\multirow{5}{*}{C} & obs. counts & 245 & 100 & 306 & 255 & 26 & 932 \\
		& \% obs. counts & 26.29 & 10.73 & 32.83 & 27.36 & 2.79 & 100.00 \\
		& Base & 20.7 & 21.9 & 21.6 & 21.1 & 19.4 & 20.7 \\
		& Data 1 & 1.8 & 2.1 & 2.3 & 3.4 & 4.9 & 2.4 \\
		& Data 2 & 1.7 & 1.9 & 1.6 & 1.7 & 1.5 & 1.6 \\
		& & & & & & & \\
		
		\multirow{4}{*}{Unclassified} & obs. counts & 126 & 120 & 48 & 26 & 9 & 329 \\
		& \% obs. counts & 38.30 & 36.47 & 14.59 & 7.90 & 2.74 & 100.00 \\
		& Base & 24.7 & 25.5 & 26.0 & 27.0 & 30.6 & 26.0 \\
		& Data 1 & 2.1 & 1.3 & 2.3 & 4.7 & 3.7 & 2.3 \\
		& Data 2 & 2.0 & 1.2 & 1.3 & 2.3 & 3.4 & 2.0 \\
		\bottomrule
	\end{tabular}
\end{table}

\section{Discussions on the limitations of the methodology}
The empirical data gathered in this study vary significantly from the standard volume-delay curve. The context-specific saturation delay and saturation speed functions and the standard BPR volume-delay function show significant variations in the residuals. The large scatter observed in the volume-delay relationship must be due to unknown external factors that have not been included in the candidate models.

\subsection{Example scenario \& possible factors} \label{sec:factors}

\begin{figure}[htbp!] 
	\centering    
	\includegraphics[width=0.7\textwidth]{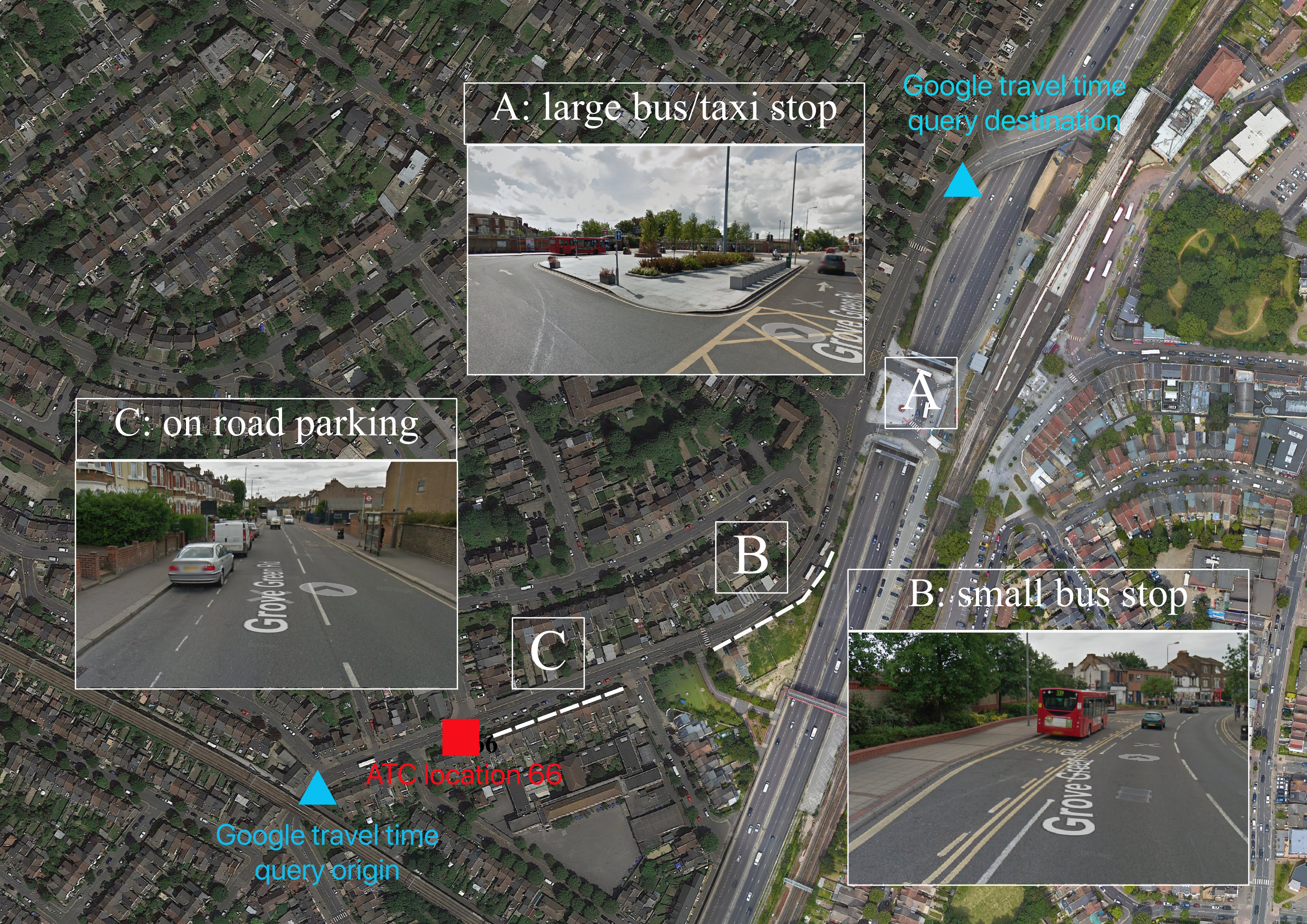}
	\caption[Satellite \& street view images of ATC 66~\citep{google_maps}]{Satellite \& Street View images of ATC 66~\citep{google_maps}}
	\label{fig:road_characters}
\end{figure}

\cref{fig:challenge_situations} (a) presents the observed volume and delay scatter plot for ATC location 66 in both directions. It is obvious that the southbound direction consistently experiences higher delay than the northbound direction, even at same volumes of traffic. In an attempt to better understand and possibly explain this distinct directional behaviour, satellite photography and street-level photography of the road is assessed. \cref{fig:road_characters} shows a satellite image of ATC location 66 obtained from Google Maps~\citep{google_maps}. The red marker gives the location of the ATC counter itself, and the two blue markers illustrate the origin/destination location used in the Google Directions API request. From this satellite imagery and street view imagery, possible explanatory factors can be identified:

\begin{enumerate}
	\item The southbound lane features road parking;
	\item The southbound lane features a large bus stop and taxi lay-by (for Leytonstone Station) and a smaller bus stop lay-by.
\end{enumerate}

The larger bus stop and taxi lay-by serving Leytonstone is a significant geometric feature that is likely to have significant impact on the southbound traffic. The second smaller bus stop lay-by and on-road parking may also have an impact, albeit to a smaller extent. Such factors may explain the exhibited differences from the location-device-informed journey data and thus permit their inclusion in an informal way.

\subsection{Challenging scenarios}

\begin{figure}[htbp!] 
	\centering    
	\includegraphics[trim={0 0 0 0},clip, width=1.0\textwidth]{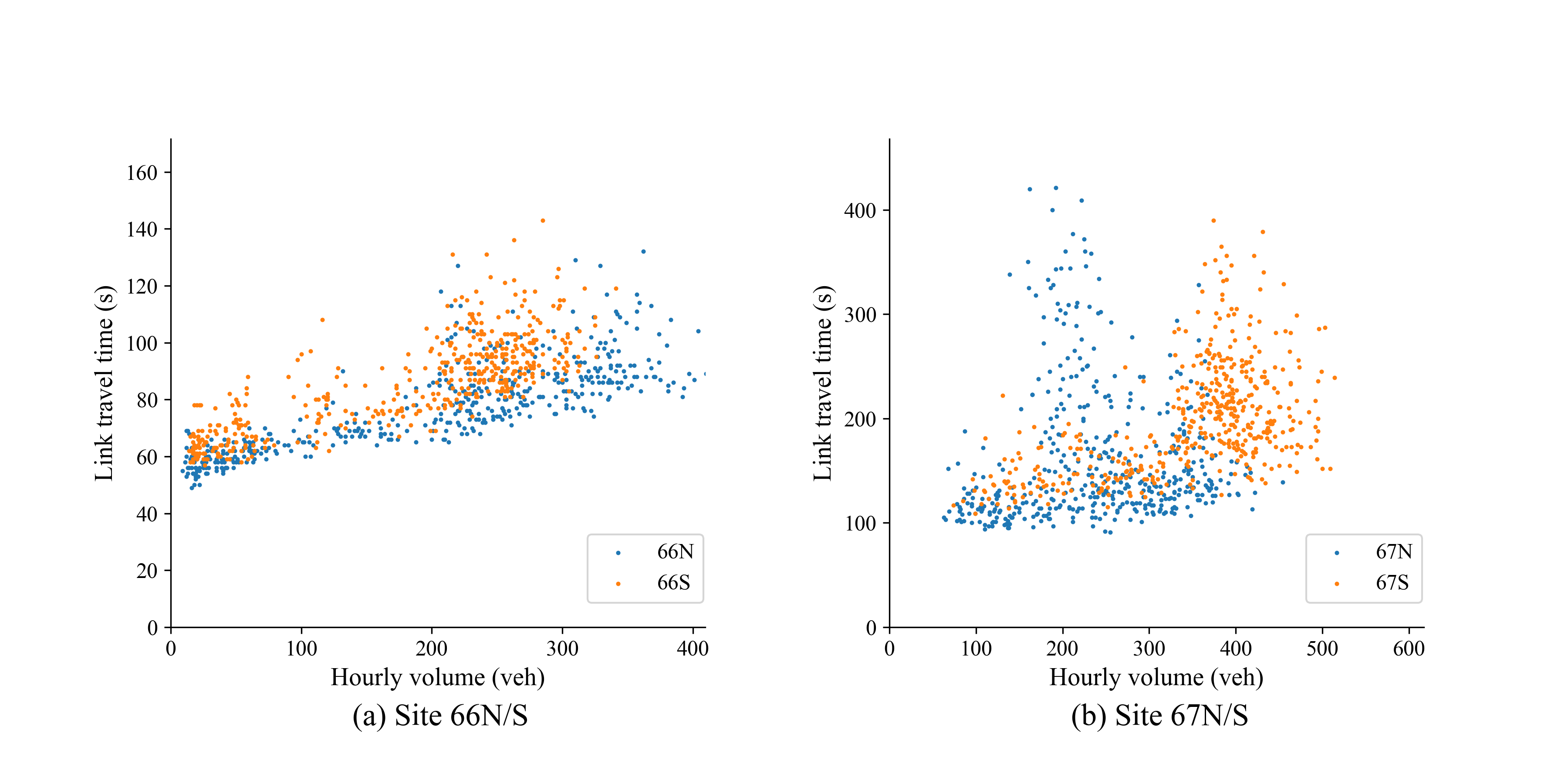}
	\caption{Challenging scenarios: volume and delay observations at ATC site 66 (a) and 67 (b)}
	\label{fig:challenge_situations}
\end{figure}

Inspections of the volume-delay scatter plots reveal a subset of roads that are not suitable to be described with the BPR-style monotone function. For example, as previously explained in \cref{sec:cleaning}, 15 sites are eventually removed from the analysis due to invalid data. Among the remaining 24 sites, many have outliers even after the data cleaning process (e.g., \cref{fig:clean}). \cref{fig:challenge_situations} (b) shows another ``abnormal" example, particularly with the northbound direction: the delay has a considerable variation when the hourly volume is around 200 vehicles, far from its capacity of about 400 vehicles per hour. This is probably resulted from the hyper-congested stage of the traffic, where the density has exceeded the optimum value and flow decreases despite continuous increase in the delay. Such phenomena cannot be correctly modelled by the monotone volume-delay function and should be dealt with using dynamic models if the hyper-congested stage is crucial in the analysis (which is usually the case for short-term predictions, such as evacuations). This is also reflected in the modeling guidance from the DfT, where the preferred choice of modeling congested urban roads in traffic assignment is to explicitly represent the junction delay and low cruise speed~\citep{DfT_webtag}. The proposed data-driven models are useful in certain scenarios, such as traffic assignment at the city-scale, where simplicity of the volume-delay curves is crucial to ensure the computational tractability of the model. However, its limitation, namely allowing unrealistic high traffic flow to be assigned to certain links, should be aware of. We plan to compare the performance of the data-driven method with the more detailed models involving junctions in our future work. The errors of using the monotonic volume-delay function in face of hyper-congested real world observations are reflected in the MAE quantification as shown in \cref{fig:mae}. Apart from these reasons, external factors could also affect the sensibility of the data, for example, a traffic collision or flooding may result in a high level of delay in a short period of time.

\subsection{Influencing factors}\label{sec:other_factors}

In \cref{sec:factors} the existence of public transport infrastructure and on-road parking were identified as being potentially explanations of the distinct southbound northbound behaviour on that road. A range of other possible factors have been identified, which may help explain the unexplained variations:

\begin{enumerate}
	
	\item \textit{Vehicle type}
	
	In this study, all vehicles are considered to be the same, despite clear distinction in the interactions between two cars compared to an interaction of a car and a lorry~\citep{vap2007investigating}. Further work is planned to assess the impact of the vehicle type. A wide variety of vehicles are observed on different road types, and such differences should be considered. The use of statistical vehicle mix sampled by road type~\citep{DfT_car_stats} was discounted for this study due to the small sample size of such statistics compared to the resolution of the data used here. The use of traffic cameras with number plate recognition and sufficient privileges to the Driver and Vehicle Licensing Agency (DVLA) database would enable the disaggregation of vehicle type at a similar spatial and temporal resolution to the Google Directions and DfT ATC data presented here. 
	
	\item \textit{Weather}
	
	Weather events may impact the journey times by altering the functionality of the vehicle, the performance of the road and/or the performance of the driver. Ongoing research focuses on combining a large dataset of Google Directions journey times with data from the UK's Met Office NIMROD precipitation dataset~\citep{met_office} to assess the relationship between these variables. At a microscopic level, it is known that an increase in precipitation increases journey times as a result of increased risk and the resulting decrease in vehicle speeds to compensate for this factor~\citep{mashros2014impact}. 
	
	\item \textit{Road incidents}
	
	Road works and road traffic collisions can lead to decreased or even zero capacity on a road link, resulting in increased saturation which impacts the journey time. Depending on the warning before such an event, the vehicle traffic may have the ability to adjust to this information, resulting in a greater distribution of traffic thus reducing the mediated travel times. Alternatively, an accident may occur without any warning to other road users committed to their route choice, resulting in longer journey time and perhaps gridlocks in the extreme cases. A method incorporating different accident and roadworks databases with Google Directions data is currently being investigated.
	
	\item \textit{Road geometry, type \& land use}
	
	Different road layouts may increase the complexity of vehicle interactions. For example, the curvature of a corner and the road surface quality will impact the speed of a vehicle. The surrounding land use also has an influence on the vehicle speeds (for example, drivers take safety precautions near a school or leisure centre). The inclusion of such factors poses many challenges, including the size and complexity of the data plus the uncertainty and variability in how drivers react to the data. In \cref{fig:road_characters}, a series of geometric factors are displayed as an attempt to explain the different behaviours on the same road in different directions. The factors shown in \cref{fig:road_characters} such as on-road parking and bus lay-bys could be quantitatively captured using computer vision and data sources such as Google Street View.
	
\end{enumerate}

\subsection{Generalisation and application}
This study proposes an efficient data collection and processing pipeline for calibrating the macroscopic volume-delay relationships for urban roads. The methodology considers the context-specific factors that might affect such relationships but are hard to quantify using conventional approaches. Apart from the challenges of using simple volume-delay curves in capturing the complex traffic behaviour, another bottleneck for the proposed methodology is the availability of the data. This limitation comes in two parts. First of all, the volume data could only be obtained for a small number of fixed locations, where the ATCs are installed and can provide continuous traffic counts. The DfT also conducts manual traffic surveys at a larger number of locations. However, these manual count data usually come at a lower frequency (e.g., annual data) and may again miss the temporal network changes that have an impact on traffic (e.g., roadworks). This bottleneck could be overcome in the near future by employing alternative data sources of traffic flow information, such as the information extracted from traffic cameras using computer vision techniques~\citep{drake2018opencv}. Secondly, the traffic speed data from Google (or other mobility data providers) are not free, which may require the cost of acquiring data to be budgeted in for large-scale studies. However, this method is expected to be cheaper and more scalable than collecting the speed/delay data from conventional methods, such as the floating car data method.

Nonetheless, the proposed methodology could still be beneficial in understanding the performance of key network locations, such as bridges or major routes, when only point-based traffic measurements are available. In addition, if more types of traffic sensors are installed in the future (e.g., radar or camera), the proposed methodology could then be applied on a larger scale.

\section{Conclusions}

By combining two disparate data sources from the ATCs and Google's traffic speed data, context-specific volume-delay curves for a range of different locations and road types have been generated for selected roads in Greater London. The derived volume-delay curves informed by real observational data have shown significant improvements in capturing individual road characteristics compared to the standard model. Apart from obtaining the site-specific knowledge, in a more practical use-case, the derived functions could now be used in the traffic assignment stage of the traditional four-step model. In some cases, the data presents clear evidence that unknown external factors, such as those listed in \cref{sec:other_factors}, have a significant impact on the traffic behavior and warrant further investigation. In these cases, both the derived functions and the standardised curve deviate significantly from empirical data and as such their use should be considered with care.

Overall, this paper has demonstrated the feasibility of linking the ATC traffic count data and Google's traffic speed data for better characterizing the volume-delay relationship across a range of sites. These data sources have longevity and are available at real-time. In the case of the real-time traffic speed data, they can be obtained from partnership with mobility service providers such as Google or its competitors (e.g., Bing, TomTom), leveraging their mature platforms and products. The methods presented in this paper may be employed over a long time horizon and at a finer temporal resolution in order to better understand the recurrent temporal and spatial trends as well as the impacts of special influencing factors, such as sporting occasions and weather events. The automation of this method over longer time horizons may lead to better explanations for various external factors influencing traffic flow behaviour and highlight areas that require investigation in order to better understand the performance of road infrastructure.

\begin{Backmatter}

\paragraph{Acknowledgments}
The authors are grateful to both Google and the DfT for access to data. We would also like to than the Engineering and Physical Sciences Research Council (EPSRC) for sponsoring this work.

\paragraph{Competing interests}
None

\paragraph{Data availability statement}
The data and code for this paper have been made open and can be accessed at https://github.com/cb-cities/volume-delay-curves.

\paragraph{Funding statement}
The project is funded by the Engineering and Physical Sciences Research Council (EPSRC) Industrial Cooperative Awards in Science \& Technology (I-CASE) studentship in collaboration with Arup. The funder had no role in study design, data collection and analysis, decision to publish, or preparation of the manuscript.

\paragraph{Ethical standards}
The research meets all ethical guidelines, including adherence to the legal requirements of the study country.

\paragraph{Author contributions}
Gerard Casey and Kenichi Soga conceptualised the project and methodology. Gerard Casey, Bingyu Zhao, Krishna Kumar did the analysis and writing with Kenichi Soga reviewing and editing. 

\bibliographystyle{apalike}
\bibliography{ref}

\end{Backmatter}

\end{document}